\documentstyle[pictex,amssymb]{article}

\newcommand{\complex}{{\mathbb{C}}}
\newcommand{\nat}{{\mathbb{N}}}

\newcommand{\real}{{\mathbb{R}}}
\newcommand{\integer}{{\mathbb{Z}}}
\newcommand{\semioplus}{\mbox{$\subset\!\!\!\!\!\!+$}} 

\newcommand{\ts}{\textstyle}

\newcommand{\grad}{\rm grad}

\renewcommand{\th}{\theta}

\newcommand{\eqref}[1]{\textup{(\ref{#1})}}
\newcommand{\bpsi}{\bar \psi}
\normalbaselines 
\begin{document}
 \pagestyle{empty}

 \rightline{IC/2002/67;\ \ {quant-ph/0207077;\ \ {}July 2002} }

 \vspace{15mm}

 \begin{center}

 \textsf{\LARGE Quantum Mechanics with Difference Operators}

 \vspace{10mm}

 V.K. Dobrev \\ Institute of Nuclear Research and Nuclear Energy, \\
         Bulgarian Academy of Sciences, 1784 Sofia, Bulgaria\\
dobrev@inrne.bas.bg\\ 
and\\
The Abdus Salam International Center for Theoretical Physics\\
P.O. Box 586, 34100 Trieste, Italy\\
dobrev@ictp.trieste.it\\[5mm]
          H.-D. Doebner
\\ Arnold Sommerfeld Institut\thanks{\,\, Now at Department of Physics}, 
TU Clausthal\\
          38678 Clausthal-Zellerfeld, Germany \\ 
asi@pt.tu-clausthal.de\\[5mm]
          R. Twarock\\ Department of Mathematics, City University\\
          Northampton Square, London EC1V 0HB\\
          r.twarock@city.ac.uk 
 \end{center}

 \vspace{10 mm}

\begin{abstract}
A formulation of quantum mechanics with additive and multiplicative ($q$-)
difference operators  instead of differential operators is studied from
first principles. Borel-quantisation on smooth configuration spaces is used
as guiding quantisation method. After a short discussion this method is
translated step-by-step to a framework based on difference operators. To
restrict the resulting plethora of possible quantisations additional
assumptions motivated by simplicity and plausibility are required.
Multiplicative difference operators and the corresponding $q$-Borel
kinematics are given on the circle and its $N$-point discretisation; the
connection to $q$-deformations of the Witt algebra is discussed. For a
``natural" choice of the $q$-kinematics a corresponding $q$-difference
evolution
equation is obtained. This study shows general difficulties for a
generalisation of a physical theory from a known one to a ``new" framework.

\end{abstract}

 \vspace{7mm} 

\noindent
{\bf Key words:} nonlinear Schr\"odinger equation, Witt algebra, 
$q$-deformation, discrete derivative.

 \newpage
\pagestyle{plain}
\setcounter{page}{1}

\section{Introduction}

\subsection{Motivations}

\noindent
There are different arguments to use 
{\it difference operators} ${\mathcal D}_{[x_1, x_2]}$ 
instead of differential operators 
acting on suitable (complex) function spaces.  
Such operators are  built out of {\it difference quotients},  
e.g. 
\begin{equation} 
D_{[x_1, x_2]} f(x) = \frac{f(x_2)-f(x_1)}{x_2-x_1}\,, \quad x_2 \not= x_1\,, x_1, x_2 \in \real \,;
\end{equation} 
for $x_2 \rightarrow x_1$ one gets the usual derivative. 
They can be generalized e.g. to different types of lattices in $\complex$. 

We mention some of the reasons: 
\begin{itemize}
\item[A.] \textbf{Fundamental Remark:} 

Heisenberg  wrote in 1930 in ``The Physical 
Principles of Quantum Theory" \cite{Heisenberg:1930}:  "...it seems necessary 
to demand that no concept enters a theory which has not been experimentally 
verified at least to the same degree of accuracy as the experiments explained 
by the theory."  
Hence one can argue that 
e.g. difference  quotients instead of differentials are  a more 
``physical" mathematical model for momenta, and that Borel sets instead of 
points are more appropriate to describe localization in physical 
space-time. 

\item[B.] \textbf{Space-Time as a lattice:} 

  There are physical reasons to 
assume that the configuration space and the  time, or both, have a lattice 
structure which is embedded into a smooth manifold. 
A physical 
theory on a given lattice can be formulated with the help of difference 
operators if e.g. a configuration space is not given but constructed from 
 a representation of a 
Lie algebra (or a deformed  Lie algebra) which contains generators interpreted 
 as position operators. It is 
reasonable in this framework to identify the spectra of these position operators, which may 
be continuous and/or discrete, as  configuration space \cite{Palev:1982,Palev:1997,Wess:1997,Wess:1998}.

\item[C.] \textbf{Quantum Symmetry:} 

If one assumes that a theory is based on an algebra which is a deformation 
in the sense of Drinfeld 
\cite{Drinfeld}, i.e. a $q$-deformation of a  Lie algebra, 
one gets realizations on  Hilbert spaces 
spanned by functions on  suitably chosen spaces. The 
generators of the algebra are given in terms of (discrete) $q$-derivatives and one 
finds, e.g., $q$-difference equations for the evolution \cite{Mrugalla}. 

\item[D.] \textbf{Numerical Methods:} 

For  approximate solutions of PDE  lattice  methods difference equations are useful. To check their 
accuracy, it is reasonable to base already the theory, which 
yields the PDE, on difference operators. 

\end{itemize}

In all these approaches one expects, that the formulation in terms of 
difference operators yields in the limiting case of  differential operators  the  
``usual" formulation. However, because of the different algebraic 
properties of difference operators this may not be the case.

\subsection{Choices for difference quotients and difference operators.} 

There are two principally different canonical 
types of difference quotients 
acting on appropriately chosen  function spaces.

We start with a function space over $\real$, $F[\real^1]$, where we have the following two options:  
\begin{enumerate} 

\item The \textbf{additive type} (based on the additive unit 
$a\in \real^1$, $a\neq 0$) 

\begin{equation}\label{add} 
D^a f(x)=\frac{f(x+a)-f(x-a)}{(x+a)-(x-a)}=
\frac{f(x+a)-f(x-a)}{2a}\,.
\end{equation} 

\item The \textbf{multiplicative type} (based on the multiplicative unit 
$q\in \real^1$, $q\neq 1$) 

\begin{equation}\label{mult} 
D^q f(x)=\frac{f(qx)-f(q^{-1}x)}{qx-q^{-1}x}=
\frac{1}{x}\frac{f(qx)-f(q^{-1}x)}{q-q^{-1}}\,.
\end{equation} 
\end{enumerate} 

These quotients can also be viewed  as operators acting on 
function spaces over  lattices in $\real^1$ of the following types:  

\begin{enumerate} 

\item The \textbf{additive type}  (uniform $a$-lattice) 

\begin{equation}\label{addLattice} 
{\mathcal L}_a := \{x_0 + j a \vert j \in \integer, x_0 \in \real 
\}\,. 
\end{equation} 

\item The \textbf{multiplicative type} ($q$-lattice, $q$ real) 

\begin{equation}\label{multLattice} 
{\mathcal L}_q := \{ x_0 q^j  \vert j \in \integer, x_0 \in \real , 
x_0\neq 0 \}\,. 
\end{equation} 

\end{enumerate} 

The above can be extended to the complex case which we need 
in order to consider functions on the unit circle $S^1$.
In particular, for $F[S^1]$ both  lattices appear as 
uniform $N$-point discretizations $S_N^1$ of $S^1$ 
depending on the parametrization of $S^1$ through $\phi\in [0,2\pi)$ or 
through $z=e^{i\phi}$. The additive 
type occurs for $\phi_j=\frac{2\pi j}{N}$:   
\begin{equation} 
S^1_N (a) := \{ \phi_j \vert \phi_j= aj,\, j=0,\ldots,N-1\} 
 \,, \quad a=\frac{2\pi}{N}\,, 
\end{equation} 
and $D^a$ acts on $F[S_N^1(a)]$.   
For $z_j\in\complex$ we get the multiplicative type: 
\begin{equation} 
S^1_N (q) := \{ z_j \vert z_j= q^j,\, j=0,\ldots,N-1\}\,, \quad q=e^{i\frac{2\pi}{N}}\,, 
\end{equation} 
and $D^q$ acts on $F[S_N^1(q)]$. Note that because of the coordinatisation, $q$ appears as a phase factor. 

{}From $q$-difference quotients we construct $q$-difference 
operators 
\begin{equation}\label{e8}
{\mathcal D}^qf(x) = \frac{g_1(q,x)f(qx)-g_2(q,x)
f(q^{-1}x)}{x(q-q^{-1})}+g_3(q,x)\
\end{equation} 
with $\lim_{q\rightarrow 1} g_i(q,x)=1,$ $i=1,2,$ 
$\lim_{q\rightarrow 1} g_3(q,x)=0.$ There is a plethora of 
possibilities, indicating that one needs additional physical 
and mathematical arguments to select a reasonable class of 
difference operators. 

In connection with the use of difference operators we introduce additive and 
multiplicative shift operators $K^a$ and 
$K^q$. The latter may depend in addition on $k\in {\integer}$,  
acting on suitably chosen (e.g. polynomial) complex functions $f(x)$ as follows: 
\begin{equation}\label{Ka} 
K^af(x)=f(x+a), \qquad K^a=exp(a\frac{d}{dx})
\end{equation} 
\begin{equation}\label{Kq} 
K_k^qf(x)=f(q^kx), \qquad K_k^q=
q^{kx\frac{d}{dx}}=exp \left(~(ln~q)~kx\frac{d}{dx}\right)
\end{equation} 
$a$- and $q$-numbers (operators) are defined as ($A$ denotes a number or an 
operator) 
\begin{equation}\label{Aa} 
[A]_a=\frac{\exp aA-\exp (-aA)}{2a},
\end{equation}
\begin{equation}\label{Aq}
[A]_q=\frac{q^A-q^{-A}}{q-q^{-1}},
\end{equation} 
with $\lim_{a\rightarrow 0}[A]_a=A$ and $\lim_{q\rightarrow 1}[A]_q=A$. The 
difference quotients (\ref{add}) and (\ref{mult}) -- for $k=1$ -- 
can be written as 
\begin{equation}\label{Da} 
D^af(x)=\left[\frac{d}{dx}\right]_af(x)=\frac{K^a-K^{-a}}{2a} 
f(x)\,,
\end{equation} 
\begin{equation}\label{Dq} 
D_{k}^qf(x)=\frac{1}{x}\left[ kx\frac{d}{dx}\right]_qf(x)=
\frac{1}{x}\frac{K^q_k-K^{-q}_k}{q-q^{-1}}f(x)\,.
\end{equation} 

The decomposition rule  is 
\begin{equation}\label{ABa} 
[A+B]_a=[A]_a \exp (\epsilon aB)+[B]_a\exp (-\epsilon aA),
\end{equation} 
\begin{equation}\label{ABq} 
[A+B]_q=[A]_qq^{\epsilon B}+[B]_qq^{-\epsilon A}, 
\end{equation} 
with $\epsilon=\pm 1$. 

\subsection{A path to $q$- and $a$- quantum mechanics}

A (non relativistic) quantum system is given through a quantisation map ${\mathcal Q}$ which maps 
a certain class of 
classical observables  under physically and mathematically motivated 
conditions  into the set ${\mathcal SA}(\mathcal{H})$  of self-adjoint 
operators  on a Hilbert 
space $\mathcal{H}$. In a first step the kinematics of the classical 
system, that is the class of generalized position and momentum observables, 
is quantised.   There are 
different methods to construct  such a quantisation map: 
the canonical quantisation for systems localized on $R^n$, the geometric 
quantisation and the Borel quantisation on smooth manifolds $M$ 
\cite{Angermann,Doebner:1990} and the 
inhomogeneous current algebra on $R^n$ \cite{Goldin:1968,Goldin:1970}. 
All these quantisations  model the momentum 
observables through a partial differential operator on  $\mathcal{H}$. In a second step,  a time dependence is introduced. 

If one is interested to formulate quantum 
mechanics, e.g. on one-dimensional manifolds, with difference operators 
$D^a$ or $D^q$,    
one has to shape  the quantisation map accordingly 
($a$- and $q$-quantisations). 
The quantised kinematics 
 is expected to yield an evolution difference equation. 
One gets a Schr\"odinger equation with difference  
operators from first principles and not by inserting difference 
operators into the usual Schr\"odinger equation. 

Because the mentioned quantisation methods are (partly) based 
on realizations of Lie algebras 
we expect that $a$- or $q$-quantizations lead to realizations which contain 
difference  quotients and shift operators. 
In particular, $D^q$ should yield results 
which are equal or similar to those from  the representation theory 
of $q$-deformed Lie algebras, including also deformed Lie brackets. 

We present in the following  paths to construct an $a$-- and 
$q$--quantisation 
involving $D^a$ and $D^q$. 
In section 2 we shortly review  
Borel quantisation \cite{Doebner:1990} on smooth manifolds with the example 
of $S^1$.  
In section  3  we formulate  
a  version for a $q$-quantisation (which is equivalent to an implicit $a$-quantisation, see 3.2.3) on  
$C^{\infty}(S^1)$ 
and $F[S_N^1(q)]$. The study contains a 
discussion (section 3.2) of different deformations of the inhomogeneous Witt algebra $W$. 
 For $F[S^1_N(q)]$ we  
calculate a family of  evolution difference equations;  in 
the ``continuous limit''  this family is larger than a family which we 
got from the 
quantisation map  for systems on $L^2(S^1 , d \Phi )$. 

This is a case study of the structure of quantum mechanics based on 
difference operators modelling  momentum observables, 
a corresponding deformation of the kinematical structure 
and the resulting dynamics.  We present 
 no numerical simulations, which are in addition  to the 
formal roots essential to get further physical insight into the model. 

\section{Short review of Borel quantisation with application to $S^1$}
\label{two}

We consider (non relativistic, point--like) systems $S$ moving and localized on a smooth manifold (with measure $\mu$) and specialize later to $S^1$. 

\subsection{The kinematics}

To model possible localization regions of $S$ on $M$ and possible 
infinitesimal movements of the regions we choose for the regions Borel sets $B$ from a 
Borel field ${\mathcal B}(M)$ and for the movements (smooth, complete)  
vector fields $X\in \mbox{Vect}_0(M)$. These two geometrical objects 
are the building blocks of the kinematics ${\widetilde{ K(M)}}$ of $S$: 
\begin{equation} 
{\widetilde{ K(M)}}=({\mathcal B}(M),\mbox{Vect}_0(M))\,.
\end{equation} 
Borel sets are displaced through $X$ by its flow $\Phi^X_\tau$ as 
$B'=\{m'\vert m'=\Phi^X_\tau (m), m\in B, X\in \mbox{Vect}_0(M), \tau \in [0,1)\}$. 
For a quantisation of ${\widetilde{ K(M)}}$ one has to construct a map ${\mathcal Q}=(Q,P)$ 
which maps the blocks in 
${\widetilde{ K(M)}}$ into the set $SA({\mathcal H})$ of self-adjoint operators on a Hilbert space ${\mathcal H}$. It is reasonable to interpret the matrix elements of $Q(B)$, i.e. $(\psi, Q(B)\psi)$, $\psi\in{\mathcal H}$, as the probability to find the system localized in $B$ in a state $\psi$. The properties of ${\mathcal B}(M)$ and further physical requirements (e.g. no internal degrees of freedom) show that $Q(B)$ acts on ${\mathcal H}$ as the characteristic function $\chi (M)$ of $M$, if ${\mathcal H}$ is realized via square integrable functions over $M$, i.e. as $L^2(M,d\mu)$. From the spectral theorem and from $Q({\mathcal B}(M))$ we infer a quantisation map for $C^{\infty}(M,\real)$ 
\begin{equation}
Q: C^{\infty}(M,\real) \rightarrow SA({\mathcal H}), \quad Q(f)\psi = f\psi\,. 
\end{equation} 
Hence we can use in the kinematics $C^{\infty}(M,\real)$ instead of ${\mathcal B}(M)$, that is the kinematics can be viewed as an infinite dimensional Lie algebra, more precisely as a semidirect sum of the abelian algebra $C^{\infty}(M,\real)$ and a subalgebra of the Lie algebra of vector fields, and we denote it as $K(M)$ (without tilde) in the following: 
\begin{equation}\label{K}
{{K(M)}} = C^{\infty}(M,\real) \semioplus \mbox{Vect}_0(M)\,.
\end{equation}

To construct the quantisation map $P$ for $\mbox{Vect}_0(M)$ we need further 
assumptions, which we will  call {\it $P$-assumptions}: 

\begin{enumerate}
\item The Lie structure of ${{K(M)}}$ is conserved. 
\item The operator $P(X)$ are -- in analogy to the canonical quantisation in $\real^n$ -- local differential operators. 
\end{enumerate}

With these $P$-assumptions we have the following result \cite{Angermann}: 
\bigskip

The $P(X)$ are differential operators of order one with respect to a 
differential structure  on the set $M\times \complex$. We characterize this structure (up to isomorphism) through  Hermitian line bundles 
$L$ over $M$  with compatible 
flat connection $\nabla$.  Wave functions are sections 
$\sigma (m)$ in the bundle and  
$L^2(M, \mu )$ can be viewed as a space of square integrable sections. 
Unitary equivalent irreducible maps 
${\mathcal Q}^{(\alpha, D)}$ 
-- quantisations -- are given by a bijective mapping onto the set 
\begin{equation}
(\alpha, D)\in \pi_1^* (M) \times \real\,.
\end{equation}
$\pi_1^* (M)$ denotes the dual of the first fundamental group of $M$, a topological quantity. $D$ is connected with the algebraic structure of $ K(M)$ and characteristic for Borel quantisation. $(\alpha, D)$ are quantum numbers in the sense of Wigner. 
${\mathcal Q}$ is labeled by these numbers, i.e. 
${\mathcal Q}^{(\alpha, D)}=(Q^{(\alpha, D)}, P^{(\alpha, D)})$ and   
one has ($m\in M$) 
\begin{eqnarray}
\label{QP} 
  Q^{(\alpha, D)}(f)\sigma (m)  
& = & f(m)\sigma (m) 
\\ 
 P^{(\alpha, D)}(X) \sigma(m)
& = & \ts \left(  -i \nabla_X^{\alpha } + 
  \left( -\frac{1}{2} i +D \right) 
   \mbox{div}_\mu g 
  \right) \sigma(m)\, ,
\end{eqnarray}
which are self-adjoint operators on a common dense set. Here, $\nabla_X^{\alpha }$  denotes the connection in the line bundle $L(M)$ over $M$ and $\mbox{div}_\mu$ denotes divergence. 
Note that the quantum number $D$ appears as a real factor in front of 
$\mbox{div}_\mu g$ and that the nontrivial topology yields the $\alpha$ 
dependence of $\nabla^{\alpha}_X$. 

\subsection{The dynamics}

States of $S$ are modeled via density matrices $W$, i.e. 
through trace class operators with $\mbox{Tr}(W)=1$. 
We introduce a time dependence for $W$ (in the Schr\"odinger picture), 
which is based on ${\mathcal Q}^{(\alpha,D)}$, through a quantum analog 
to the classical relation between time derivatives $\frac{d}{dt}$ of time dependent 
functions $f(m(t))$ and momenta, i.e. for $M=\real^1$ 
\begin{equation} 
\frac{d}{dt} f(x(t)) \sim p \nabla f\,.
\end{equation}
One can show \cite{Doebner:1995} that (in the Schr\"odinger picture) 
one has the following relation for expectation values 
($\mbox{Exp}_W({\bf A})=Tr(W{\bf A})$): 
\begin{equation}
 \frac{d}{dt}\mbox{Tr}(W(t) Q^{(\alpha,D)}(f))=\mbox{Tr}(W(t) P^{(\alpha,D)}\left(X_{\grad f} \right))\,, \forall f \in C^\infty (M,\real)
\end{equation}
This is a restriction for the evolution of $W(t)$. For pure states  
it implies,  under the condition that pure states evolve into pure states 
\cite{Doebner:1995}, 
the following generalized version of the first 
Ehrenfest relation 
\begin{equation}\label{Ehr}\label{25}
 \frac{d}{dt}
  \left\langle \sigma(t), Q^{(\alpha, D)}(f) \sigma(t) \right\rangle 
  = 
  \left\langle \sigma(t), P^{(\alpha, D)}\left(X_{\grad f} \right) 
\sigma(t) 
  \right\rangle, \quad \forall f\in C^{\infty}(M,\real)
\end{equation}
with a scalar product $\langle .,. \rangle$ in $L^2(M,d\mu )$.

\subsection{The  kinematical algebra ${{ K(S^1)}}$ and a family of evolution equations}

We consider now an application of Borel quantization to the case that the configuration space is $S^1$. 
$S^1$ is topologically nontrivial with $\pi _1^*(S^1)=[0, 2\pi )$, and we denote elements in 
$\pi _1^*(S^1)$ as $\alpha$. The flat line bundles over $S^1$ are trivial, 
the vector fields 
are $X=X(\phi)\frac{d}{d\phi}\in \mbox{Vect}_0(S^1)$ and  
the Hilbert 
space is $L^2(S^1, d\varphi)$. 
In these coordinates ${{ K(S_1)}}$ is given by the  generators 
\begin{eqnarray} 
 Q^{(\alpha, D)}(f) \psi (\phi) 
& = & f(\phi)\psi(\phi) \label{27}
\\ 
  P^{(\alpha, D)}(X) \psi (\phi) 
& = & \ts \left(  -i  X(\phi)\frac{d}{d\phi}+ 
  \left( -\frac{1}{2} i +D \right) 
  \left( \frac{dX(\phi)}{d\phi} \right) 
  +\alpha X(\phi)\right) \psi(\phi)\,. \label{28}
\end{eqnarray} 
To analyse the structure of ${K(S_1)}$  
we use a Fourier transform ${\mathcal F}$ with $z=e^{i\phi}$: 
\begin{equation}\label{e28} 
\label{efX} 
\begin{array}{rcl} 
\displaystyle 
f(\phi)=  \hat{f}(z(i\phi)) &  = & 
   \displaystyle \sum_{n=-\infty}^{\infty} f_n  z^n \\ 
  X(\phi)= \hat{X}(z(i\phi)) & = & 
   \displaystyle \sum_{n=-\infty}^{\infty}   X_n z^n, 
\end{array} 
\end{equation} 
$f_n ={\bar f}_{-n}$, $X_n={\bar X}_{-n}$. For the 
${\mathcal F}$-transformed quantum kinematics 
we find 
\begin{equation} 
\label{eIa} \label{32}
\begin{array}{rl} 
\displaystyle 
 Q^{(\alpha , D)}(f) = & \displaystyle \sum_{n=-\infty}^{\infty} f_n z^n \\ 
 P^{(\alpha , D)}(X) = & \displaystyle \sum_{n=-\infty}^{\infty} X_n  z^n 
         \left( z \frac{d}{dz} +{\frac{n}{2}} 
         + \alpha +iDn \right)\,. 
\end{array} 
\end{equation} 
With the operators 
\begin{equation} 
\label{gen} 
\begin{array}{rl} 
\displaystyle 
 T_n = & \, z^n \\ 
 L_n^\alpha = & 
 \displaystyle 
 z^n \left( z \frac{d}{dz} +{\frac{n}{2}} + \alpha 
 \right) 
\end{array} 
\glossary{$T_n$} 
\glossary{$L_n^\th$} 
\end{equation} 
\eqref{eIa} can be expressed as 
\begin{equation} 
\label{ops} \label{e32}
\begin{array}{rl} 
\displaystyle 
 Q^{(\alpha , D)}(f) = & \displaystyle \sum_{n=-\infty}^{\infty} f_n T_n \\ 
 P^{(\alpha , D)}(X) = & \displaystyle \sum_{n=-\infty}^{\infty} X_n 
     \left( L_n^\alpha +iDnT_n \right)\,. 
\end{array} 
\end{equation} 
The generators $T_n$ are an Abelian Lie algebra which we denote as $T$,  
and for fixed $\alpha\in [0, 2\pi)$ the 
$L_n \equiv L_n^{\alpha}$ fulfill the commutation 
relations 
\begin{eqnarray} 
 \left\lbrack  T_m,T_n \right\rbrack & = & 0 \nonumber\\ 
\label{eCRGen} 
 \left\lbrack  L_n,T_m \right\rbrack & = & mT_{m+n} \\ 
 \left\lbrack  L_m,L_n \right\rbrack & = & (n-m)L_{m+n} \nonumber
\end{eqnarray} 
and span an inhomogenisation of the  Witt algebra $W$ through $T$. 
This gives the  
algebraic structure of ${{K_z(S^1)}}$ \cite{Doebner:1990}, where we use the index $z$ to indicate that it is given in terms of the variable $z$ as opposed to the angle variable $\phi$. We have from 
\eqref{eCRGen} (we have dropped the upper index $(\alpha , D)$ for convenience)
$$
[Q(f), Q(g)]=0,\quad [P(X),Q(f)]=-iQ(Xf), \quad [P(X), P(Y)]=-iP([X,Y]).
$$
Now we introduce the time dependence for pure states 
$\psi(\phi)\in L^2(S^1 , d\phi )$ and we 
evaluate 
the restriction (\ref{Ehr}) 
with $X_{\grad f}=f'(\phi)\frac{d}{d\phi}$ 
($' \equiv \frac{d}{d\phi}$):    
\begin{equation}
 \frac{d}{dt}
  \left( \psi, f \psi \right) 
  = 
  \left( \psi , P^{(\alpha , D)}\left(f'\frac{d}{d\phi}\right)\psi \right), 
\quad \forall f\in  C^{\infty}(S^1,\real).
\end{equation}
This implies 
a generalized continuity equation of Fokker-Planck type for 
$\rho=\bpsi\psi$: 
\begin{equation}\label{35}\label{e35}
\dot \rho
 =\frac{i}{2} ({\bar \psi} \psi''-{\bar \psi}'' \psi)
     +D  \rho'' -\alpha \rho'=-(j_0^\alpha )'+D\rho'' ,
\label{rho}
\end{equation}
where 
$$ j_0^\alpha = 
 \frac{i}{2} ({\bar \psi}' \psi-{\bar \psi} \psi')+\alpha\rho 
$$
corresponds for vanishing $\alpha$ to the usual quantum mechanical current 
density on $S^1$. 

This can be derived also by other methods based on $Q^{(\alpha , D)}$ 
\cite{Doebner:1992}, \cite{Doebner:1993a}, \cite{Doebner:1994}.
We use this information in (\ref{35}) for a general Ansatz for a Schr\"odinger equation of the type
$$
i \partial_t \psi=H\psi+G[\bar \psi, \psi] \psi
$$
in which $H$ is a linear operator and $G[\bar \psi, \psi]$ 
can be written (formally) as a nonlinear complex function $G[\bar \psi, \psi]=G_1[\bar \psi, \psi]+iG_2[\bar \psi, \psi]$ 
depending on $\bar \psi, \psi $, their derivatives and explicitly on 
$\phi$ and $t$. Hence $G$ acts as a multiplication operator. 
This Ansatz leads to a family ${\mathcal F}_P$ of Schr\"odinger  
equations \cite{Twarock:1997a}, \cite{Twarock:Phd} 
on $L^2(S^1 , d\phi)$ with $G_2$ enforced by~(\ref{rho}) 
\begin{equation} \label{SE}\label{36} 
 i \partial_t 
 \psi 
 = -\frac{1}{2} \frac{d^2}{d\phi^2} \psi 
   -i\alpha \frac{d}{d\phi} \psi 
   +i\frac{D}{2 \rho} 
    \left( \frac{d^2}{d\phi^2} \rho \right) \psi 
   + G_1[\bar \psi, \psi]\psi.
\end{equation} 
The real part $G_1$ cannot be determined by Borel quantization. Hence a set of (natural) assumptions for $G_1$ motivated 
by  the form of the imaginary part $G_2$, has been introduced 
\cite{Doebner:1992}, \cite{Doebner:1994}:
\begin{enumerate} 
\label{realpart} 
\item 
$G_1$ is proportional to $D$, i.~e.\ vanishing for $D=0$. 
\item $G_1$ is a rational function with derivatives no higher than second order and occurring in the numerator only. 

\item $G_1$ is complex homogeneous of order zero, i.~e.\ $G_1\lbrack \alpha\psi, {\bar \alpha} {\bar \psi} \rbrack = G_1\lbrack \psi, {\bar \psi} \rbrack$ for all $\alpha\in\complex$. 

\end{enumerate} 
These assumptions  restrict $ G_1$ in the family ${\mathcal F}_P$ to the  \textbf{Doebner-Goldin family } (DG-family) ${\mathcal F}_{DG}$ \cite{Doebner:1992} on $S^1$: 

\begin{equation}\label{DG} 
 G_1\lbrack \bar \psi, \psi\rbrack:= 
 D_1 \frac{\left(j_0^\alpha\right)'}{\rho} 
 + D_2 \frac{\rho''}{\rho} 
 + D_3 \frac{\left(j_0^\alpha\right)^2}{\rho^2} 
 + D_4 \frac{\left( j_0^\alpha \rho' \right)}{\rho^2} 
 + D_5 \frac{\left( \rho' \right)^2}{\rho^2} 
\end{equation} 
with free real parameters $D_k$, $k=1,\ldots,5$. 

\section{An $a$- and $q$- quantisation of the kinematical algebra $ {{ K(S^1)}}$}

\subsection{Strategy  and  remarks}

After the review of Borel quantisation and their application to the configuration space $S^1$ 
we follow our programme  to construct an analogous realisation of 
${{ K(S^1)}}$ with difference instead of differential operators.  For this we need  guiding principles.  As 
in section 2 we  use  $L^2(S^1, d\Phi)$ or the Hilbert space of 
functions over the lattice  $S^1_N(q)$ with elements $\psi=(\psi(0),\ldots,\psi(N-1))$, $\psi(j)=\psi_j \in \complex$, with inner product 
\begin{equation}\label{inner}\label{hilbert}
(\psi,\phi) = c(\psi,\phi) \sum_{j=0}^{N-1} {\bar \psi_j} \phi_j\,,
\end{equation}
where $c(\psi,\phi)$ is a suitable normalisation. For the operator $Q(f)$ we use the corresponding  multiplication operator. If applied to difference or shift operators  the P-assumptions in section 2 fail  to determine $P(X)$ \footnote{To realise $P(X)$ as differential operator we have introduced a 
differentiable structure on $M\times\complex$ via a complex line bundle 
over M; the algebraic properties of ${{K(M)}}$ restrict the 
order of $P(X)$ to one.} and we need further  principles. 

We start  from the results for ${{K(S^1)}}$ in section 2.   Our strategy for $P(X)$ is to replace 
first order differential operators in ${{K_z(S^1)}}$ through ``first order'' 
difference operators or shift operators. Another option in the plethora of possibilities is a deformation 
of the Lie bracket.  The 
following assumptions -- called $d$-assumptions\footnote{$d$ for deformations} -- will be implemented: 

\begin{enumerate}
\item The difference (shift)  operators  should be chosen such that  the 
corresponding realisation of ${{ K(S^1)}}$ is again an algebra with a Lie bracket 
or a deformed Lie bracket. We cannot expect a realisation isomorphic to 
${{ K(S^1)}}$  but a  more general deformation of  ${{ K(S^1)}}$, e.g. a non-commutative 
and/or non-cocommutative Hopf algebra or a higher order Lie algebra. 

\item  In the limit $a \rightarrow 0$ or $q\rightarrow  1$ the realisation  
should give the ``old'' result. 

\item  If the difference (shift) operators (see e.g. (\ref{e8})) depend on 
additional objects (like constants, functions) the number of these  
objects should be as small as possible and they require a physical interpretation. 

\item  We model the influence of the topology of $S^1$ through the term 
proportional to  $\alpha $  in ${{K_z  (S^1)}}$\footnote{In the discrete case $a$ and $q$ thus ``feel'' the topology of $S^1$.}.  

\item The Fokker-Planck type equation for the time dependence of the positional density should have a reasonable interpretation.

\end{enumerate}

\noindent
We add a more technical remark:
\smallskip

\noindent
The multiplicative and additive difference operators are constructed  from shift operators $K^a$  and $K^q_k$. The algebra ${{K_z  (S^1)}}$ acts  on complex polynomial functions  $\hat{f}(z)$ on $\{z\vert z=e^{i\Phi}\}$. To use difference operators in  ${{K_z  (S^1)}}$ this space must be invariant under the shift operators. For $K^q_1$ we have 
\begin{equation}\label{K1}\label{38}
K_1^q\hat{f}(z)=\hat{f}(qz), ~~ q=e^{i\Phi_0}.
\end{equation}
Hence it is necessary to use $q$ which is a phase  here. From 
\begin{equation}
\hat{f}(qe^{i\Phi})=\hat{f}(e^{i\Phi+i\Phi_0})
\end{equation}
we infer that $K^q_1$ acts as an implicit additive shift operator $I^a$ defined as 
\begin{equation}\label{40}
{I}^a\hat{f}(e^{i\Phi})=\hat{f}(e^{i\Phi+a}),
~~a=i\Phi_0.
\end{equation}
An explicit additive shift operator $K^a$  with 
\begin{equation}
K^a\hat{f}(z)=\hat{f}(z+a)\,,\quad a=z_0\,,
\end{equation}
does not exist  in this space. This is one of the reasons to use $q$-quantisations. Furthermore, we denote 
\begin{equation}\label{dot}
z\frac{d}{dz} = N_z\,.
\end{equation}

\subsection{Deformations of the inhomogeneous Witt algebra}

\subsubsection{The multiplicative setting: a natural Ansatz}

With our $d$-assumptions  we replace in the generators $L_n^\alpha$ in ${{K_z(S^1)}}$ 
\begin{equation}\label{41}
L_n^\alpha = z^n A_n^\alpha ,\quad A_n^\alpha = N_z +\frac{n}{2} + \alpha\,,
\end{equation}
the differential through $q$-difference operators and the generators $T_n$ of $T$ (formally) through ${\mathcal T}_n$ with 
\begin{equation}\label{42}
{\mathcal T}_n = T_n\,.
\end{equation}
A comparison with (\ref{28}) shows that 
the origin of the factor $z^n$ in (\ref{41}) is the function $X(\phi)$ which corresponds to the position observable. The differential $-i \frac{d}{d\phi}$ in (\ref{28}) implies $z\frac{d}{dz}$ and also the term $\frac{n}{2}$. We assume to handle $\alpha$ in the same way as $\frac{n}{2}$. Hence, we try the Ansatz 
\begin{equation}
L_n^\alpha \mapsto {\mathcal L}_n^\alpha= z^n [A_n^\alpha]_q, \quad q=\exp i\phi_0\,.
\end{equation}
The ${\mathcal L}_n^\alpha$ should close under Lie brackets, but this is not the case
which indicates a more general Ansatz. Indeed, with an additional running parameter ${k\in \integer}$ and  $q=\exp i\phi_0 $ fixed, the  
\begin{equation}\label{44}
{\mathcal L}_{n,k}^{\alpha} = z^n \frac{[kA_n^{\alpha}]_q}{[k]_q}\,, \quad k, {n\in\integer},\, \alpha\in \real\,
\end{equation}
form a Lie algebra, a $q$-deformed Witt algebra $W_q$. The commutators are for $j_1 \not = j_2$  
\begin{equation}\label{45}\label{47}
\begin{array}{rcl} 
\displaystyle 
 \left\lbrack  
{\mathcal L}_{m,j_1}^{\alpha}, {\mathcal L}_{n,j_2}^{\alpha}
 \right\rbrack 
 & = & 
 \displaystyle 
  \frac{\big\lbrack  j_1 \frac{n}{2}-j_2\frac{m}{2} 
      \big\rbrack \big\lbrack j_1+j_2\big\rbrack}{\lbrack 
      j_1\rbrack \lbrack j_2\rbrack}
{\mathcal L}_{m+n,j_1+j_2}^{\alpha}
  \\ 
 & & 
 \displaystyle 
 + \frac{\big\lbrack  j_1 \frac{n}{2}+j_2\frac{m}{2} 
      \big\rbrack \big\lbrack j_2-j_1\big\rbrack}{\lbrack 
      j_1\rbrack \lbrack j_2\rbrack} 
  {\mathcal L}_{m+n,j_2-j_1}^{\alpha}
\end{array} 
\end{equation} 
and for $j_1 = j_2$  
\begin{equation}\label{46}
\begin{array}{rcl} 
\displaystyle 
 \left\lbrack  
 {\mathcal L}_{m,j}^{\alpha},  {\mathcal L}_{n,j}^{\alpha}
 \right\rbrack
  & = & 
  \displaystyle 
  \frac{\left\lbrack  j (n-m)\right\rbrack 
  \lbrack 2j\rbrack}{\lbrack j\rbrack^2}
{\mathcal L}_{m+n,2j}^{\alpha}
\end{array} 
\end{equation} 
  
Concerning $d$-assumption 2 we have that 
\begin{equation}
\lim_{q\rightarrow 1}  {\mathcal L}_{n,k}^{\alpha}
 = L_n^\alpha\,.
\end{equation}
$ {\mathcal L}_{n,k}^{\alpha}$ is a difference operator (cf. (8)) which is symmetric (see section 3.2.3) and bounded (because of $\vert q\vert =1$) in the Fourier space: 
\begin{equation}
 {\mathcal L}_{n,k}^{\alpha}
 {\hat f} (z) = z^n 
\frac{e^{i\phi_0 (\frac{n}{2}+\alpha)} {\hat f}(e^{ik \phi_0}z)-
e^{-i\phi_0 (\frac{n}{2}+\alpha)} {\hat f}(e^{-ik \phi_0}z)}{2i\sin (\phi_0 k)}
\end{equation}
and the additional parameter $k\in\integer$ is a (presumably minimal) method to enforce a closure. 
The commutator (\ref{45}), (\ref{46}) allows to restrict the parameter $k$ to sets 
\begin{equation}\label{C}
C(p,k_0):=\{k\vert k= p k_0 \nu\,, \nu, p\in \nat, k_0 \in \integer \mbox{ fixed }  \}\,.
\end{equation}
In the following, we use $k_0=1$. 

The coupling between ${\mathcal T}_n$ and ${\mathcal L}_{n,k}^{\alpha}$ is not semidirect. From (\ref{42}), (\ref{44}) we get 
\begin{equation}\label{49}\label{52}
{\mathcal L}_{m,k}^{\alpha}{\mathcal T}_n  =  {\mathcal T}_{-n}  {\mathcal L}_{m+2n,k}^{\alpha}
\end{equation}
i.e. a quadratic coupling.  The $\{{\mathcal L}_{n,k}^{\alpha},{\mathcal T}_n\}$ with $q=\exp i \phi_0$ span\footnote{If one considers $\alpha\in [0,2\pi)$ not as a fixed but as a running parameter, the basis $\{{\mathcal L}_{n,k}^{\alpha}
, {\mathcal T}_n \vert \alpha\in \real\,, k, n\in \integer\}$ is for $q$ fixed again a quadratic Lie algebra\cite{Twarock:1997a}. }
 the $q$-deformed inhomogeneous Witt algebra $W_q$. For $q\rightarrow 1$ we find the (undeformed) inhomogeneous Witt  algebra. 

\subsubsection{The Multiplicative setting: other attempts and a Hopf structure}

The above ``natural'' $q$-deformation of the inhomogeneous Witt algebra is one example in the plethora of possible  deformations. We mention other attempts obtained from a different background, e.g.,  
\cite{Aizawa:1991,Chaichian:1991,Oh:1994,Kemmoku:1996,Sato:1996} 
and present some ``generalisations'' of the above ``natural'' construction. We use in this section the Abelian Lie algebra $K$ spanned by the $q$-shift operators $K_k^q$, $k\in\integer$: 
\begin{equation} 
[K_k^q,K_{k'}^q]=0
\end{equation}
with $\lim_{q\rightarrow 1} K_k^q=1$ and with 
\begin{eqnarray}\label{51}
K_k^q {\mathcal T}_n & = & q^{kn} {\mathcal T}_n K_k^q\\
K_k^q  {\mathcal L}_{n,l}^{\alpha} & = & q^{kn} {\mathcal L}_{n,l}^{\alpha}
 K_k^q
\end{eqnarray}
\begin{itemize}
\item[a)] 
One can enlarge the basis $\{ {\mathcal L}_{n,k}^{\alpha}\}$ through the basis of the algebra $K$. 
The augmented basis $\{ {\mathcal L}_{n,k}^{\alpha}, {K}_m^{q}\}$ leads to a quadratic Lie algebra which is a noncommutative and non co-commutative Hopf algebra -- Hopf-q-Witt algebra -- ${\mathcal H}W_q$ with coproduct $\Delta$, counit $\epsilon$ and antipode $\gamma$ given as follows: 
\begin{equation}\label{Hopfstructure}
\begin{array}{rl}
\Delta({\mathcal L}_{m,j}^{\alpha})={\mathcal L}_{m,j}^{\alpha}
\otimes K_m                        +K_m \otimes {\mathcal L}_{m,j}^{\alpha}
 &
\Delta(K_l)=K_l\otimes K_l \\\
\epsilon({\mathcal L}_{m,j}^{\alpha})=0   &
\epsilon(K_l)=1 \\\
\gamma({\mathcal L}_{m,j}^{\alpha}
)=-K_m^{-1}{\mathcal L}_{m,j}^{\alpha}
 K_m^{-1} &
\gamma({K_l})=K_l^{-1}
\end{array}
\end{equation}
This construction is of interest because $W_q$ is cocommutative: the price for a nontrivial Hopf algebra is the enlargement through an infinite dimensional Lie algebra. An inhomogeneous Hopf-$q$-Witt algebra with basis 
$\{ {\mathcal L}_{n,k}^{\alpha}
, {K}_m^{q}, {\mathcal T}_l\}$ follows from (\ref{49}), (\ref{51}). A restriction to $\{ {\mathcal L}_{n,k}^{\alpha}
, {\mathcal T}_l\}$ yields the same results as in 3.2.1. and also the same dynamics as in section 4. 

\item[b)] 
To change $\{ {\mathcal L}_{n,k}^{\alpha}
, {\mathcal T}_l\}$ and the resulting dynamics one option is to deform not only the differential quotients but -- generalizing our strategy -- also the position operators ${T}_n$     
(i.e. the $Q(f)$). Their deformation forms a non-commutative algebra, which is related to some kind of non-commutative geometry. We try this option and multiply each ${\mathcal T}_n$ with a shift operator $K_{\delta}^q$ with fixed $\delta\in \integer$, 
\begin{equation}
T_n \mapsto {\widetilde {\mathcal T}}_n = {\mathcal T}_n K_\delta^q
\end{equation}
Relation (\ref{51}) implies 
\begin{equation}
{\widetilde {\mathcal T}}_m{\widetilde {\mathcal T}}_n =q^{\delta(n-m)} {\widetilde {\mathcal T}}_n {\widetilde {\mathcal T}}_m
\end{equation}
The ${\widetilde {\mathcal T}}_n$ couple to $\{ {\mathcal L}_{n,k}^{\alpha} \}$  according to 
\begin{equation}
q^{-\delta n}{\widetilde {\mathcal T}}_m{\mathcal L}_{n,j}^{\alpha+m}
 = {\mathcal L}_{n,j}^{\alpha} {\widetilde {\mathcal T}}_{m}
\end{equation}
which reduces to (\ref{49}) for $\delta =0$, i.e. $K_0^q=1$. The deformed algebra 
$\{ {\mathcal L}_{n,k}^{\alpha}, {\widetilde {\mathcal T}}_m\}$ is quadratic. As mentioned in 3.2.1 the 
factor $z^n$ is the result of a position observable and one should replace $z^n={\mathcal T}_n$ now by 
${\widetilde {\mathcal T}}_n$. This gives 
${\widetilde {\mathcal L}_{n,k}^{\alpha} } = {\mathcal L}_{n,k}^{\alpha} K_\delta^q$. 

\item[c)] 
Another reasonable Ansatz is to deform the three terms in $A_n^\alpha$ separately and multiply them with different shift operators: 
\begin{equation}
\label{57}
L_n^\alpha \mapsto {\mathcal A}_n^{\alpha,I} = z^n \left(
q^{\lambda_1} \frac{\left[\alpha_1 N_z \right]_q}{[\alpha_1]_q}K_{\beta_1}^q +
q^{\lambda_2} \frac{\left[\alpha_2 z \frac{n}{2}\right]_q}{[\alpha_2]_q}K_{\beta_2}^q +
q^{\lambda_3} \frac{\left[\alpha_3 z \alpha \right]_q}{[\alpha_3]_q}K_{\beta_3}^q 
\right)
\end{equation}
with $I$ denoting the nine real parameters $\lambda_i$, $\alpha_i$, $\beta_i$, $i=1,2,3$, depending on $q$, $\alpha$, $n$ and on additional parameters like $k$. The $d$-assumption  
\begin{equation}
\lim_{q\rightarrow 1} {\mathcal A}_n^{\alpha,I} = L_n^\alpha
\end{equation}
is fulfilled. The ${\mathcal A}_n^{\alpha,I}$ are a special case of ${\mathcal L}_{n,k}^{\alpha}$ for  
\begin{equation}
\begin{array}{ccc}
\lambda_1 = k (\frac{n}{2} + \alpha), & \lambda_2 = k \alpha, & \lambda_3 =- k \frac{n}{2}\\
\beta_1 =0, & \beta_2 = \beta_3 = -k, & \alpha_1=\alpha_2 = \alpha_3 = k \,.
\end{array}
\end{equation}
It is not possible to close $\{{\mathcal A}_n^{\alpha,I}\}$ if not at least one running parameter is introduced; the structure of the commutators and the coupling to ${ \mathcal T}_n$ or ${\widetilde {\mathcal T}}_n$ will be discussed in another context. 
\item[d)]
In the above constructions, the Lie bracket is not deformed. A $q$-commutator 
\begin{equation}
\label{58}
[A,B]_q := q^{r(A,B)} AB - q^{s(A,B)} BA
\end{equation}
with real valued functions $r$, $s$ depending on $A$ and $B$ gives another freedom. We consider an example for a $q$-deformation of $W$ with a deformed Lie bracket (\ref{58}) and deformed generators (\ref{57}) which contain only two fixed real parameters $a$ and $b$ and no running parameter, like $k$ in (\ref{44}). For simplicity, we treat the case $\alpha=0$. If we choose in ${\mathcal A}_n^{0{I}}$
 the set ${I}$ as 
\begin{equation}
\begin{array}{ccc}
I: & \lambda_1 = n (a+1)b, & \lambda_2 = nab\\
& \alpha_1=-\beta_1 =b, & \alpha_2=-\beta_2 = 2b
\end{array}
\end{equation}
and a $q$-commutator for $ [ {\mathcal A}_n^{0{I}}, {\mathcal A}_m^{0{I}}]_q$ with 
\begin{equation}
 r(n,m)=-2mb, \quad s(n,m)=-2nb
\end{equation}
then it closes for ${\mathcal A}_n^{0{I}}$:  
\begin{equation}
\label{59}
[ {\mathcal A}_n^{0{I}}, {\mathcal A}_m^{0{I}}]_q =  \frac{(q^{-2bn}-q^{-2bm})}{[b]} {\mathcal A}_{n+m}^{0{I}}\,. 
\end{equation}
For $q\rightarrow 1$ the Witt algebra $W$ is obtained. The price which one has to pay in order to have no running parameter $k\in\integer$ but only two fixed parameters is a $q$-commutator. 
\end{itemize}

\subsubsection{An implicit additive setting}

Concerning deformations of ${{K_z(S^1)}}$ through additive shift operators we refer to the fact (see (\ref{38}),(\ref{40})) that a multiplicative shift $K^q {\hat f}(z) = {\hat f}(qz)$, $\vert q \vert =1$, can be viewed as an implicit additive shift 
${I}^a {\hat f}(z(i\phi)) = {\hat f}(z(i\phi+a))$, $a=i\phi_0$. 
Such a shift reproduces the results in 3.2.1. Accordingly, we use as a deformation of $L_n^\alpha$ 
\begin{equation}
\label{60}
L_n^\alpha \mapsto {\mathcal M}_{n,k}^{\alpha}
\end{equation}
with  
\begin{equation}
\label{61}
{\mathcal M}_{n,k}^{\alpha} = z^n \frac{[k A_n^I]_a}{[k]_a}\,, \qquad a=i\phi_0
\end{equation}
which gives 
\begin{equation}
\label{62}
 {\mathcal M}_{n,k}^{\alpha}   ={\mathcal L}_{n,k}^{\alpha}\,.
\end{equation}
For the deformation types in 3.2.2 similar results hold. 
If one wants non-implicit additive shifts $K^a f(\phi) = f(\phi + \phi_0)$ one has to work with ${{K_\phi (S^1)}}$, that is the kinematical algebra parametrized by $\phi$ instead of $z$. 

\subsubsection{On Witt algebra deformations}

Five deformation types of the Witt algebra $W$ and its inhomogenisation 
have been analysed. A deformation $W_q$  is given in section 3.2.1 via a Lie algebra with
a running integer parameter $k$. The inhomogenisation  of $W_q$  is quadratic. A
nontrivial Hopf algebra ${\mathcal H}W_q$  and  a corresponding inhomogenisation   
are obtained if $W_q$ is
enlarged through an Abelian algebra of $q$-shifts (see 3.2.2. - a). For the inhomogeneous Hopf-$q$-Witt algebra  we
get the same dynamics as for the inhomogeneous $W_q$ as we demonstrate later. If one deforms not only $W$ but also the
commutative algebra with basis  $\{ {\mathcal T}_n\}$ one finds non-commutative 
position
operators and hence some link to a noncommutative geometry (3.2.2-b).
Generalisations of $W_q$ were given in 3.2.2 - c. A deformation of the Lie
bracket $[.,.] \rightarrow [.,.]_q$ in 3.2.2 - leads for a special family
of generalised $W_q$ to a Lie algebra without a running parameter but with two fixed
parameters. Implicit additive deformations are equivalent to $q$-deformations
(3.2.3). For non-implicit additive deformations one has to use 
${{K_\phi(S^1)}}$. In
the limit $q\rightarrow 1$ one gets for all types $W$, or respectively, its inhomogenisation. 
There are further
types obtained from a different background. We mention  [15 - 19] among
other attempts.

If one analyses the different types of $q$-deformations with respect to how the 
$d$-assumptions are ``fulfilled'', it seems that (44) in 3.2.1 is a preferred
choice  (``natural'' 
Ansatz):\footnote{Because the $d$-assumptions  are not ``sharp'', also other types can be viewed as ``natural''.}  
$W=\{L_n^\alpha\}$  is a Lie
algebra. The running parameter  $k$ in the deformation $\{ {\mathcal L}_{n,k}^{\alpha} \}$ behaves such that it can be interpreted
as ``internal degree of freedom'' which couples to  
$W$. It becomes apparent in the evolution equation  (see 4.), e.g. for the $N$-point discretisation of $S^1$ with $q = exp~ i {2\pi\over{N}}$, and may be interpreted as follows: A representation of a (non Abelian) Lie algebra in terms of differentials is 
local in the sense that differentials are 
 local objects because they depend only on a small neighbourhood of the point at which they are evaluated.\footnote{For a deformed Lie bracket the situation may change.} 
A difference operator, on the contrary, when evaluated at $x$ depends  
on $x$ and $q^k x$ or $x+ka$ for some $k\in\integer$. Hence a representation of a ``deformed'' Lie algebra via difference 
operators depends on $k$, which we view as internal degree of freedom, and thus each point $x$ depends, via the discrete derivative, on this one-dimensional ``internal'' space $S_{int}$, which is an indication for how coarse-grained the difference operator is with respect to the underlying discrete space.

There are further peculiarities: A differential quotient is related via its
inverse to a corresponding integral which is used to construct 
$L^2(S^1,d\Phi)$. 
For a $q$-difference quotient the Jackson integral \cite{Jackson} is
the appropriate notion. One can apply such an integral for a function space over $S^1$
 to model a $q$-inner product; however, this is not convenient here. 

 The choice of an inner product is relevant if
one wants the position and momentum operators to be symmetric (or even
self-adjoint). 
In the Fourier space the ${\mathcal L}_{n,k}^{\alpha}$ and ${\mathcal T}_n$ 
are symmetric, also the $Q(f)$
and $P(X)$, and for the shift operator
$$
(K_k^q)^*=K_k^{-q}
$$
holds.

\subsection{Deformations of the kinematical algebra ${{K_z(S^1)}}$}

The deformation of the inhomogeneous Witt algebra (\ref{47}), (\ref{52}) leads almost directly to a deformation ${{K_z(S^1)}}_q$ of ${{K_z(S^1)}}$; for the other deformation types in 3.2.2  (except 3.2.2.- d), the results are the same or similar. 

Insert ${\mathcal L}_{n,k}^{\alpha}$ and  ${\mathcal T}_n$ for $L_n^\alpha$ and $T^n$, respectively, in (\ref{ops}). The origin of the term $iDn$ is $\frac{d}{d\phi} X(\phi) =iz \frac{d}{dz} {\hat X}(z)$ in (\ref{28}). 
According to our strategy, we replace 
\begin{equation}
\label{67}
\frac{d}{d\phi} \mapsto i [N_z]_q
\end{equation}
in (\ref{28}), thus  
\begin{equation}
\label{68}
D \frac{d}{d\phi} X(\phi) \mapsto iD \sum_n [N_z]_q X_n z^n = \sum_n iD [n]_q X_n z^n\,.
\end{equation}
With (\ref{42}), (\ref{44}), (\ref{68}) the deformation $K_z(S^1)_q$ is (the upper index $(\alpha,D)$ is dropped)
\begin{equation}\label{opsDef}
\label{69}
\begin{array}{rcl}
Q_q (f) & = & \sum_n f_n {\mathcal T}_n\\[0.2cm]
P_q^k (X) & = & \sum_n X_n ({\mathcal L}_n^{\alpha,k} + i [n]_q D {\mathcal T}_n)\,, k\in S_{int}
\end{array}
\end{equation}
For any $k$, i.e. for any point $k$ in the internal space $S_{int}$, we get a momentum operator $P_q^k$. The $Q_q$ are independent of $k$. $P_q^k$ applied to ${\hat f}(e^{i\phi})$ gives a linear combination of ${\hat f}(e^{i(\phi\pm k \phi_0)})$
and ${\hat f}(e^{i\phi})$ which depends on $n$ and is multiplied by $z^n$. 

For $S_N^1$ and the corresponding Hilbert space (\ref{hilbert}) one can see directly the (``non local'') action: For $\phi_j=\frac{2\pi}{N}j$, $j=0,\ldots,N-1$, $q=\exp i\frac{2\pi}{N}$, the momentum $P_q^k$ at the point $j$ depends on $j$ and $j\pm k$. This implies $k\in C(1,1)$.

\section{Dynamics from a $q$-deformed kinematical algebra ${K_z(S^1)_q}$}

\subsection{Strategy}

In section 2 a time dependence for pure states in the Schr\"odinger representation was introduced through a generalized Ehrenfest relation (\ref{25}) with a quantisation $Q^{(\alpha, D)}$ of the kinematical algebra ${{K(M)}}$ as an input. We use the same construction here with ${{K_z(S^1)}}_q$ as input. 
Hence the time dependence of $\psi_t$ is restricted through 
\begin{equation}
\label{70}\label{75} \label{e75}
\frac{d}{dt} (\psi_t, Q_q(f) \psi_t) = (\psi_t , P_q^k (\grad f) \psi_t)\,, \quad \forall f\in C^{\infty}(S^1)\,.
\end{equation}
We have not used the option to substitute the time derivative $\frac{d}{dt}$ through a reasonable time difference operator because we want to analyze the right hand side of (\ref{70}), that is the space depending part.
For a discretisation using a $t$-difference operator see \cite{Twarock:Phd}. For $M=\real^n$, $n=1,2$, a $q$-deformation of the Schr\"odinger equation through a group theoretical approach has been considered in \cite{Mrugalla,ADD}. 

Relation (\ref{70}) depends on $k\in S_{int}$ but the dynamics do not couple different points $k$ in the internal space. The derivation of (\ref{25}) is based on the operators $Q_q$ and $P_q^k$; the algebraic relations of this kinematics, which couples different points of $S_{int}$, are not used in the derivation.\footnote[7]{It would be interesting to introduce coupling in $S_{int}$ such that one gets for $q\rightarrow 1$ the non-deformed result.}
For any $k\in S_{int}$ (\ref{70}) yields a corresponding evolution equation. Note that the internal coordinate $k$ couples only to the ``external'' configuration space. 

As we have mentioned before we focus our consideration on $S_N^1$ and a Hilbert space with elements $\psi_t =(\psi_t(0),\ldots,\psi_t(N-1))$, $\psi_t(l\pm N) = \psi_t(l)$, with inner product (\ref{hilbert}). As shown in 1.2 we have to choose $q=\exp i \frac{2\pi}{N}$. The Fourier transformation is 
\begin{equation}
\label{71}
\psi_t(l)=\sum_{k=1}^{N-1} \psi_{kl} z_l^k,\quad z_l = \exp i \frac{2\pi}{N} l\,.
\end{equation}
The time dependence in the discrete case is restricted by (\ref{75}) with the corresponding inner product and valid for all $f=(f_0,\ldots,f_{N-1})$, $f_j\in \real$. In Fourier space, the situation for $S^1$ and $S^1_N$ differ through the Fourier sums and $q$. However, we treat formally both cases together and specify later.

\subsection{The deformed generalized $q$-Ehrenfest relation for $S_N^1$}

On the left hand side of (\ref{70}) we insert the expression for $Q_q(f)$ in (\ref{opsDef}) and on the right hand side, we use  
$\grad f(\phi)=\frac{d}{d\phi} f(\phi) = i N_z {\hat f} (z)$. Following (\ref{68}) this implies with the Fourier transform (\ref{e28}) and (\ref{71}) 
\begin{equation}
\grad f = i \sum_{n} [n]_q f_n z^n 
\end{equation}
and 
\begin{equation}
\label{71a}
P_q^k (\grad f)=\sum_{n} i [n]_q f_n ({\mathcal L}_{n,k}^{\alpha} +iD [n]_q {\mathcal T}_n)\,.
\end{equation}
Because (\ref{75}) holds for all $f$ we have for the $t$-dependence of the probability density $\rho_t={\bar \psi}_t \psi_t$ 
\begin{eqnarray}
{\dot \rho}_t & =&  \sum_{l,j} {\bar \psi}_t(l) z^{-l} \psi_t(j) z^j c_{lj} \label{exp}\\
c_{l,j} & = & i [l-j]_q \left( 
\frac{[k(j+\frac{j}{2}(l-j)+\alpha]_q}{[k]_q}+iD[l-j]_q 
\right)\label{73}
\end{eqnarray}
This is a $q$-version of the Fokker Planck type equation (\ref{35}):  
\begin{eqnarray}
{\dot \rho}_t & = & - \frac{\partial}{\partial \phi} (I_0^\alpha - D \frac{\partial}{\partial \phi} \rho_t)\label{dotrho}\label{e81}\\
I_0^\alpha & = & \frac{i}{2}(\frac{\partial}{\partial \phi} {\bar \psi} \psi - {\bar \psi} \frac{\partial}{\partial \phi} 
\psi) + \alpha \rho \label{current}\label{e82}
\end{eqnarray}
To show this, replace in (\ref{dotrho}) and (\ref{current}) $\frac{\partial}{\partial \phi}$ by $i[N_z]_q$ and construct a deformed current ${I}_0^{\alpha,k}$ such that 
\begin{equation}
\label{74}
{\dot \rho}_t = -i [N_z]_q ({ I}_0^{\alpha,k} - iD[N_z]_q \rho_t) 
\end{equation}
yields (\ref{exp}). Such a current exists and is given by ($q=\exp i \Phi_0$) 
\begin{equation}\label{e84}
\begin{array}{rcl}
 {I}_0^{\alpha,k} & = &  
\frac{1}{[k]_q} 
\left[
\left( 
[\frac{k}{2}(N_z+\alpha)]_q \psi 
\right)
\left(
K_{\frac{k}{2}}^q \exp (i\frac{k}{2} \alpha \Phi_0) {\bar \psi} 
\right)\right.\\
&&\left.
- 
\left(
K_{\frac{k}{2}}^q \exp (-i \frac{k}{2} \alpha \Phi_0) \psi
\right)
\left(
 [\frac{k}{2} (N_z - \alpha)]_q {\bar \psi} 
\right) 
\right]
\end{array}
\end{equation} 
For $q\to 1$, $I_0^{\alpha,k}$ gives $j_0^\alpha$ in (\ref{e35}) and (formally) 
$I_0^{\alpha,k}= - I_0^{-\alpha,k}$ holds. 

The shift operators $K_{\pm \frac{k}{2}}^q$ act on ${\bar \psi}(l)$ and on $\psi (l)$ in both terms in 
${I}_0^{\alpha,k}$ and yield $\psi (l\pm \frac{k}{2})$. In the case of $S_N^1$ we assure that $l\pm \frac{k}{2}$ corresponds to a lattice point in $S_N^1$. This implies (see (\ref{C})) 
 \begin{equation}
\label{76}\label{e85}
k\in C(p,1)\,, p \mbox{ even }
\end{equation}
thus $k=p \nu$, $\nu \in \nat$. 
Hence, the interaction between next neighbours via the discrete internal space $S_{int}$ depends on the choice of $p$ and $\nu$. 

\subsection{The Evolution Equation}

The $q$-Fokker-Planck type equation (\ref{74}) restricts the evolution of $\psi_t$. In the undeformed case the Ansatz (note the multiplication by ${\bar \psi}$) 
\begin{equation}
\label{77}\label{e87}
i {\dot \psi}_t{\bar \psi_t} = (H \psi_t + G({\bar \psi}_t,\psi_t) \psi_t){\bar \psi}_t
\end{equation}
implies in the usual Fokker-Planck type equation (\ref{35}) a certain form for $H$ and $G_2$. An analogous method fails in (\ref{e84}) because the $K^q$ act on ${\bar \psi}$ and on $\psi$. One possibility to overcome this is an Ansatz for $i {\dot \psi}_t{\bar \psi_t}$ which contains $N_L$ suitably chosen shifts (we skip the index  $t$ in $\psi$), and a splitting of the linear part into $N_L$ units, each associated with a different shift $S^l$ of ${\bar \psi}$, as follows:  
\begin{equation}
\label{78}
i {\dot \psi}{\bar \psi} = \sum_{l=1}^{N_l} (H^l_q({\bar \psi},\psi) \psi) (S^l {\bar \psi})+
 (G_q ({\bar \psi},\psi) \psi )( R {\bar \psi})
\end{equation}
$S^l$ and $R$ are shifts which may be expressed as $(a_1 K_{k_1}^q + a_2 K_{k_2}^q)$ with $a_1$, $a_2\in\complex$, $k_1$, $k_2\in C(p,1)$; $H_q^l$ for each $l$ is a complex linear difference operator, $G_q$ is a (formal) nonlinear multiplication operator (see the analogy to 2.3). A corresponding expression follows for $i{\dot {\bar \psi}} \psi$. The Ansatz was used in \cite{Twarock:1997a}. We indicate the calculation: 

Split $S^l$, $R$, $\psi$, ${\bar \psi}$, $H_q^l$, $G_q$ in real (index 1) and imaginary (index 2) parts. Insert from (\ref{e87}) ${\dot {\bar \psi}}\psi$ and ${\bar \psi}{\dot \psi}$ in the left hand side of (\ref{74}) and get 
 a relation between $H_{q,i}^l$, $G_{q,i}$, $S^l_{i}$ and 
$R_{i}$, $i=1,2$. We collect the linear operators on $\psi$ and on ${\bar \psi}$ in $F_{L}$ to determine $H_q^l$ and $S^l$ 
and the nonlinear terms in $F_{NL}$ which give some information on $G_q({\bar \psi},\psi)$ and $R$: 
\begin{equation}
\label{79}
i {\dot \psi} {\bar \psi} = F_L +F_{NL}
\end{equation}
Then a straightforward calculation shows that 
\begin{equation}
\label{81}
\begin{array}{rcl}
F_L & = & \left([N_z]_q \frac{[\frac{k}{2}N_z]_q}{[k]_q} \psi \right) \left(S_{q,1} {\bar \psi}\right)\\
 && - i\frac{[\frac{k}{2}\alpha]_q}{[k]_q} \left\{
(\exp (-i\frac{k}{2}\alpha \Phi_0) i[N_z]_q K_{\frac{-k}{2}}^q \psi) (S_{q,1} {\bar \psi}) 
\right.\\ 
&& \left.  + ( \exp (i\frac{k}{2}\alpha \Phi_0)i[N_z]_q K_{\frac{k}{2}}^q \psi)
( K_{-k}^q S_{q,1} {\bar \psi}) \right\}
\end{array}
\end{equation}
with 
\begin{equation}
\label{82}
S_{q,1} = \frac12 \left(K_{\frac{k}{2}+1}^q + K_{\frac{k}{2}-1}^q \right)\,, \quad S_{q,2}=0\,, \quad a_l=2\,.
\end{equation}
As $q\rightarrow 1$ ($\Phi_0 \rightarrow 0$) one has by construction 
\begin{equation}
\label{83}
\lim_{q\rightarrow 1} F_L=\left[(\frac12 N_z^2 +\alpha N_z)\psi \right]{\bar \psi} \,. 
\end{equation}
For a suitable Ansatz for $F_{NL}$ remember that in the non-deformed case $G=G_1 +i G_2$ where the term $G_2$ was given through (\ref{35}); $G_1$ could be zero (the corresponding DG-equation for $M=\real^3$ was given in \cite{Doebner:1992,Doebner:1994}). Hence we assume here the same, i.e. 
\begin{equation}
\label{84}
G_q^I=iG_{q,2}^I\, \mbox{ with }\quad G_{q,1}^I =0\,. 
\end{equation}
Because $G_{q2}^I$ is a real multiplication operator, this implies after a straightforward calculation using (\ref{78}), its complex conjugate, and (\ref{74}): 
\begin{equation}
\label{85}\label{e94}
G_{q,2}^I 2 \mbox{ Re} (\psi R {\bar \psi}) = C_q^I - D [N_z]_q^2 \rho
\end{equation}
with 
\begin{equation}
\begin{array}{rcl}
\label{86}
C_q^I & = &  \frac{1}{2i} \frac{1}{[k]_q} \left[ 
-
\left(
[N_z]_q K_{\frac{k}{2}}^q \psi 
\right) \left(
[\frac{k}{2} N_z]_q (K_1^q+K_{-1}^q) {\bar \psi} 
\right)
+ \right.\\
& & \left.
\left(
[\frac{k}{2} N_z]_q (K_1^q+K_{-1}^q) {\psi}
\right) \left(
[N_z]_q K_{\frac{k}{2}}^q {\bar \psi} \right)
\right] \,. 
\end{array}
\end{equation}
$C_q^I$ vanishes for $q \rightarrow 1$ and 
\begin{equation}
\label{87}\label{e96}
\lim_{q\rightarrow 1} G_{q,2}^I = - \frac{D}{2} \frac{N_z^2\rho}{\rho}
\end{equation}
holds and gives the nonlinear imaginary term in (\ref{36}). With 
$F_{NL} = i(G_{q,2}^I (  {\bar \psi},\psi)\psi)( R {\bar \psi})$ we arrive at a nonlinear difference equation for ${\dot \psi}$
\begin{equation}
\label{88}
i {\dot \psi} = \psi^{-1} (F_L + iG_{q,2}^I (  {\bar \psi},\psi)\psi R {\bar \psi}) 
\end{equation}
In general, admitting also $G_{q,1}^I \not = 0$, one finds again for $q \rightarrow 1$ the imaginary term  (\ref{e96}). In this setting, there are $G_q$ and $R$ such that the term 
$\frac{D \rho''}{2\rho}$ in the DG-family (\ref{36}) is reproduced in the limit (see Theorem 7.15 in \cite{Twarock:Phd}). However, the choice $R_{1}=0$ or $R_{2}=0$ (see \cite{Twarock:1997a,Twarock:Phd}) gives equations not in the DG family, e.g. a nonlinear term proportional to 
\begin{equation}
 \frac{{\bar \psi}'' \psi' - {\bar \psi}'\psi''}{{\bar \psi}' \psi - {\bar \psi}\psi'}\,. 
\end{equation}
For $S_N^1$ we find $N$ coupled nonlinear difference equations. 
To illustrate the time dependence of ${\dot \psi}(l)$, $l=0,\ldots,N-1$, $\psi(l\pm N)=\psi(l)$, we consider the case $G^I$,   $\alpha = 0$. Hence the second part in (\ref{81}) vanishes. Evaluating (\ref{88}) with (\ref{81})  and (\ref{e94}) we see that the detailed form of the family of evolution equations depends on $\nu$ (because $k_{\pm}=1\pm \frac{k}{2}$ with $k= 2 \nu$, $\nu\in \nat $ (see (\ref{e85}) ), $D\in\real$ and on the shift $R$:
\begin{equation}
\label{89}
\begin{array}{rcl}
i {\dot \psi} (l) & = & \displaystyle \frac{1}{{\bar \psi}(l)} \frac{1}{\sin \frac{2\pi}{N}\sin \frac{2\pi k}{N}} 
\left\{
-1/8(2\psi(l+k_+)-\psi(l+k_-)-\psi(l-k_-))
\right.\\[0.2cm]
&&\displaystyle \left. ({\bar \psi}(l+k_+)+{\bar \psi}(l+k_-)) + 
\frac{1}{2i} \frac{1}{2\mbox{ Re}(\psi (l)R {\bar \psi}(l))} \left( 
(\psi(l+k_+)-\psi(l+k_-)\right.\right.
\\[0.2cm]
&& \displaystyle \left. \left.
+\psi(l-k_-)- \psi(l+k_+))({\bar \psi}(l+k_+)-{\bar \psi}(l+k_-))+cc.\right) R {\bar \psi}(l)
\right\}\\[0.2cm]
&& \displaystyle - \frac{1}{\mbox{ Re}(\psi (l)R {\bar \psi}(l))} \frac{D}{2} 
(\psi(l+2){\bar \psi}(l+2)- 2 \psi(l){\bar \psi}(l)+\psi(l-2){\bar \psi}(l-2))\,.
\end{array}
\end{equation}
A convenient choice for the shift operator is $R=1$. There  exists an evolution equation for any even $k\in\nat$. (\ref{89}) shows which points interact with a given point $l$. We give no numerical study on this evolution equation in this report. 

\section{Summary and Outlook}

This study, which also reviews some earlier work \cite{Twarock:1997a}, \cite{Twarock:1999a}, shows some possibilities to develop from first principles a
framework for quantum mechanics on a configuration space $M$, such that momentum observables are represented
through difference operators whereas position observables are quantised  through multiplication operators. 

We have applied Borel quantisation. This method is based on the kinematical algebra ${{ K(M)}}$ spanned by usual position observables; momentum observables are quantized with $P$-assumptions through certain differential operators. The dynamics is introduced via a generalized Ehrenfest relation and yields for pure states a
family  ${\mathcal F}_P$  of nonlinear differential equations which contain as a ``natural'' subfamily the DG family  ${\mathcal F}_{DG}$.

For the development of a framework involving difference operators we start with ${{ K(M)}}$  -- not from a kinematics ${\widetilde{K(M)}}$ with Borel sets -- and replace the quantized vector fields  through difference operators. This is a highly
non unique procedure. A minimal condition we require is that we get the results from Borel quantisation  in a suitable limit, but one needs further assumptions, which we have called  $d$-assumptions (in analogy to the $P$-assumptions). The guiding principles were  simplicity, plausibility and some ``physical feeling''. We gave the results for difference operators of the multiplicative type -- so called $q$-deformations of ${{ K(M)}}$ in the case of $M= S^1$  and  its discretisation $S_N^1$.

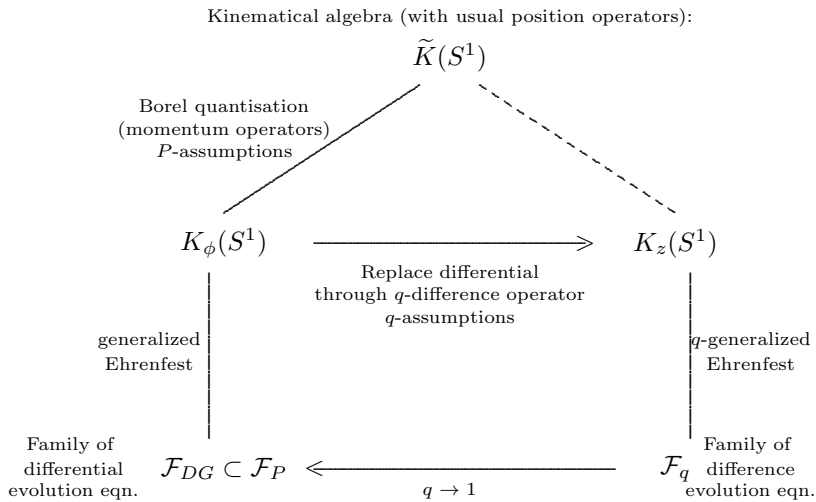
\beginpicture
 \setshadegrid span <.5truept>
 \setcoordinatesystem units <1cm,1cm> point at 0 0
 \setplotarea x from -4.5 to 4.5, y from -2.0 to 6.4

\plot -0.4 5.1 -3.0 3.4 /

 \plot 2.2 0  -1.8 0 /
 \plot 3.2 0.4  3.2 2.6 /
 \plot -3.2 0.4  -3.2 2.6 /
 \plot  1.8 3.0  -1.8 3.0 /

\put {$<$} at -1.8 0
\put {$>$} at 1.8 3.0 

\setdashes <1mm> \setlinear 
\plot 0.4 5.1 3.0 3.4 /

	 \put {${\mathcal F}_{DG}\subset {\mathcal F}_P$} at -3 0 
   \put {${\mathcal F}_q$} at 3.0 0
	 \put {$K_{\phi}(S^1)$} at -3.0 3.0 
   \put {$K_{z}(S^1)$} at 3.0 3.0
   \put {${{\widetilde K(S^1)}}$} at 0 5.5

\put {\scriptsize Kinematical algebra (with usual position operators):} at 0 6.0
\put {\scriptsize Borel quantisation} at -3.0 4.8
\put {\scriptsize (momentum operators)} at -3.0 4.5
\put {\scriptsize $P$-assumptions} at -3.0 4.2

\put {\scriptsize Replace differential} at 0 2.6
\put {\scriptsize through $q$-difference operator} at 0 2.3
\put {\scriptsize $q$-assumptions} at 0 2.0

\put {\scriptsize generalized} at -4.0 1.7
\put {\scriptsize Ehrenfest} at -4.0 1.4

\put {\scriptsize $q$-generalized} at 4.0 1.7
\put {\scriptsize Ehrenfest} at 4.0 1.4

\put {\scriptsize $q \rightarrow 1$} at 0 -0.3

\put {\scriptsize Family of } at -5.0 0.3
\put {\scriptsize differential} at -5.0 0
\put {\scriptsize evolution eqn.} at -5.0 -0.3

\put {\scriptsize Family of } at 4.0 0.3
\put {\scriptsize difference} at 4.0 0
\put {\scriptsize evolution eqn.} at 4.0 -0.3
\endpicture

The quantisation of ${{K(S^1  )}}$      is connected with a representation of the Witt algebra $W$. Hence our
construction yields certain $q$-deformations of $W$ as an interesting by-product of our study. The family  
${\mathcal F}_{DG}\subset {\mathcal F}_P$  can be 
obtained from the large family ${\mathcal F}_P$ in the limit $q \rightarrow  1$. There are ``natural'' members in 
${\mathcal F}_P$    which are not connected with ${\mathcal F}_{DG}$ in this 
limit.  We have not tried to realise the dotted line in Fig. 1.

The generalization of an existing physical theory requires the input of 
extra information to pave a path into a new structure - the right choice 
of such information is difficult; here, it was formulated  in terms of 
the $d$-assumptions. The formulation of quantum
gauge field theory in the setting of noncommutative geometry faces similar 
problems.

The step-by-step  procedure to implement difference operators may be compared with walking in a marsh jumping from one stable
looking blade of grass to the next.  For a  motivation of quantum mechanics with difference operators  a deeper
understanding and possibly a new view on the structure of our  space--time could give some necessary information. 

Finally we remark that nonlinear DG-equations have recently found attention in string theory \cite{Mavromatos:2001} and we hope that this study of discretizations of DG-equations over the configuration space $S^1$ from first principles will also be relevant in this context.

\section*{Acknowledgements}

Some of the results of this paper have been presented at the 33. Symposium on Mathematical Physics in Torun (Poland) in 2001. One of the authors (HDD) is grateful to the organizers of the conference for the warm hospitality. 
VKD would like to thank the Alexander von Humboldt foundation for financial support. 


\newcommand{\noopsort}[1]{} \newcommand{\printfirst}[2]{#1}
  \newcommand{\singleletter}[1]{#1} \newcommand{\switchargs}[2]{#2#1}

\end{document}